\documentclass[runningheads]{llncs}
\usepackage{tablefootnote}
\usepackage{amssymb}
\usepackage{amsmath}
\usepackage{graphics}
\usepackage{graphicx}
\usepackage{epsfig}
\usepackage{epstopdf}
\usepackage{mathrsfs}
\usepackage{cite}
\usepackage[mathscr]{euscript}
\usepackage{nameref}

\newcommand{\sig}{\gamma}
\newcommand{\bee}{\mathcal{B}}
\newcommand{\bl}{\beta_{\rm L}}
\newcommand{\bu}{\beta_{\rm U}}

\DeclareMathOperator*{\argmin}{arg\,min}

\newcommand{\perv}{{\rm PI}}

\newcommand{\vs}{\vspace{-1mm}}

\newcommand{\cee}{\mathcal{C}}

\newcommand{\gee}{\mathcal{G}}

\newcommand{\pset}{\ensuremath{\mathcal{P}}}

\newcommand{\Lnash}{\ensuremath{\mathcal{L}^{\rm nf}}}
\newcommand{\Lopt}{\ensuremath{\mathcal{L}^*}}

\newcommand{\Lnf}{\mathcal{L}^{\rm nf}}

\begin{document}
\title{Providing slowdown information to improve selfish routing \thanks{This work is supported in part by NSF Grant ECCS-2013779.}}
%
%
\author{Philip N. Brown\inst{1}\orcidID{0000-0003-3953-0503}}
\authorrunning{P. Brown}
%
\institute{University of Colorado Colorado Springs, CO, USA\\
\email{philip.brown@uccs.edu}}
\maketitle              
\begin{abstract}
Recent research in the social sciences has identified situations in which small changes in the way that information is provided to consumers can have large aggregate effects on behavior.
This has been promoted in popular media in areas of public health and wellness, but its application to other areas has not been broadly studied.
This paper presents a simple model which expresses the effect of providing commuters with carefully-curated information regarding aggregate traffic ``slowdowns" on the various roads in a transportation network.
Much of the work on providing information to commuters focuses specifically on travel-time information.
However, the model in the present paper allows a system planner to provide slowdown information as well; that is, commuters are additionally told how much slower each route is as compared to its uncongested state.
We show that providing this additional information can improve equilibrium routing efficiency when compared to the case when commuters are only given information about travel time, but that these improvements in congestion are not universal.
That is, transportation networks exist on which any provision of slowdown information can harm equilibrium congestion.
In addition, this paper illuminates a deep connection between the effects of commuter slowdown-sensitivity and the study of marginal-cost pricing and altruism in congestion games.

\end{abstract}

\section{Introduction}

Today, in the age of the internet of things, big data, and pervasive computing, engineered systems are becoming increasingly interconnected with the human populations that they serve.
System-level performance is directly affected by the choices of human users, customers, and adversaries, and it has long been recognized that self-interested behavior by users may lead to gross system inefficiencies~\cite{Pigou1920,Roughgarden2005}.
Thus, both the importance and the feasibility of influencing human behavior in intelligent ways are increasing simultaneously.

A popular modeling framework to study methods of influencing selfish behavior in engineered systems is the \emph{nonatomic congestion game}, which is often used to model urban transportation networks~\cite{Beckmann1956}.
This model was one of the first to admit straightforward characterizations of the \emph{price of anarchy,} a popular measure of the social cost of selfish behavior~\cite{Roughgarden2005}.
It is also a natural setting to study various modified models of human decision-making; examples include altruism~\cite{Chen2014}, pessimism~\cite{Meir2015b}, risk-aversion~\cite{Lianeas2016}, and various generalizations~\cite{Kleer2017}.
More relevant to this paper, it has proved fertile ground for the study of various methods of influencing user behavior: taxation~\cite{Cole2003,Brown2017a,Colini-Baldeschi2017,Bonifaci2011}, autonomously-controlled traffic~\cite{Biyik2021,Li2021}, and traffic information systems~\cite{Cheng2016,Acemoglu2018,Massicot2019} have all been investigated in this context.
The majority of this literature depends on the usual game-theoretic assumption that users are expected utility maximizers.

In parallel, an increasingly popular conceptual framework for studying the practice of influencing human behavior is that of \emph{behavioral economics}. This framework grew out of empirical studies which showed that humans tend to deviate from the behaviors prescribed by expected utility maximization, and that these deviations manifest themselves in predictable ways~\cite{Tversky1974}.
If humans are predictably irrational, it seems reasonable that engineers might attempt to exploit this predictability to influence behavior.
This idea has led to the concept of the ``nudge" --- the popular name given to the idea that making slight modifications to a person's environment can have a profound effect on their behavior~\cite{Thaler2008}.

This paper draws inspiration from the concept of the nudge, and asks if drivers in transportation networks could perhaps be influenced by providing simple pieces of information that are carefully designed to exploit behavioral biases in their decision-making process.
As a simple preliminary model for studying such questions, this paper begins with the typical assumption that drivers prefer low travel times over high travel times.
Some drivers, however, additionally dislike the idea of choosing a route that exhibits higher travel time than its uncongested free-flow travel time.
That is, some drivers are slowdown-averse.
A system planner, knowing about the existence (but perhaps not the magnitude) of this slowdown-aversion, may wish to exploit it to improve aggregate congestion, and provides a signal informing drivers of the presence and magnitude of slowdowns on various routes.
Note that unlike much other work on traffic information systems, we do not require ordinary drivers to perform any type of Bayesian inference; rather, our model captures a phenomenon in which each driver acts in response to her ``gut feel'' about how much she dislikes traffic slowdowns.

The main question of this paper is this: when can a planner be certain that providing slowdown information to drivers will improve congestion?
In Theorem~\ref{thm:one}, we show that networks exist for which any slowdown information provided to drivers increases equilibrium delays, and thus on these networks, the planner's only reasonable course of action is to prove no slowdown signal at all.
Nonetheless, Theorem~\ref{thm:one} also shows that on parallel networks, slowdown information can always reduce equilibrium delays.
Perhaps unintuitively, these results are tightly connected with the literature on marginal-cost pricing and altruistic behavior in transportation networks~\cite{Chen2014,Brown2020a}.
Indeed, this condition can be stated as a sufficient condition for improvements due to slowdown signaling: If altruism increases equilibrium delays on some network, then slowdown signaling also increases equilibrium delays on that network.

Following this, Theorem~\ref{thm:aggressive} essentially asks the question ``is it optimal for planners to provide accurate slowdown information to drivers?"
Alternatively, what is the optimal slowdown signal with respect to minimizing equilibrium congestion?
Here, we consider an example setting in which the planner has significant levels of information about the population's slowdown preferences, as well as some minimal information about aggregate demand.
In this setting, Theorem~\ref{thm:aggressive} shows that planners with access to this additional information optimally \emph{over-state} the severity of network slowdowns.
That is, a planner's optimal course of action may well be to deliberately provide misleading information to drivers.

Lastly, the paper closes with a note on the implications that this work has for the broader study of heterogeneous nonatomic congestion games.
Here, it is shown that many of the well-studied forms of heterogeneous congestion games appearing in the literature belong to a particularly well-behaved class of games known as \emph{weighted potential games}.
That is, the Nash equilbria of these games can be characterized as the maximizers of a single global convex potential function.
From an analytical point of view, the ramifications of this are broad: the existing work on these types of games depends on \emph{ad hoc} characterizations of equilibrium behavior, and the existence of a potential function may unify these approaches.

\section{Model and Performance Metrics}

\subsection{Routing Game}
Consider a network routing problem for a network $(V,E)$ comprised of vertex set $V$ and edge set $E$.
We call a source/destination vertex pair $(\sigma^c,t^c)\in (V\times V)$ a \emph{commodity}, and the set of all such commodities $\cee$. 
For each commodity $c\in\cee$, there is a mass of traffic $r^c>0$ that needs to be routed from $\sigma^c$ to $t^c$. 
We write $\pset^c\subset 2^E$ to denote the set of \emph{paths} available to traffic in commodity $c$, where each path $p\in\pset^c$ consists of a set of edges connecting $\sigma^c$ to $t^c$. 
Let $\pset = \cup\left\{\pset^c\right\}$.
A network is called \emph{symmetric} if there is exactly one commodity: $\cee=\{c\}$, i.e., all traffic routes from a common source $\sigma$ to a common destination $t$ using a common path set $\pset$.
A network is called a \emph{parallel} network if all commodities share a single source-destination pair and all paths are disjoint; i.e., for all paths $p,p'\in\pset$, $p\cap p'=\emptyset$.
Note that a parallel network need not be symmetric; although all traffic must share a common source and destination, the path sets $\pset^c$ available to traffic from different commodities may differ.

We write $f^c_p\geq 0$ to denote the mass of traffic from commodity $c$ using path $p$, and $f_p := \sum_{c\in\cee}f_p^c$.
A \emph{feasible flow} $f\in\mathbb{R}^{|\pset|}$ is an assignment of traffic to various paths such that for each $c$, $\sum_{p\in\pset^c}f^c_p = r^c$ and $\sum_{c\in\cee}r^c=r$.

Given a flow $f$, the flow on edge $e$ is given by $f_e = \sum_{p:e\in p}f_p$.
To characterize transit delay as a function of traffic flow, each edge $e\in E$ is associated with a specific latency function $\ell_e:[0,r]\rightarrow[0,\infty)$; $\ell_e(f_e)$ denotes the delay experienced by users of edge $e$ when the edge flow is $f_e$.
We adopt the standard assumptions that each latency function is nondecreasing, convex, and continuously differentiable.
We measure the cost of a flow $f$ by the \emph{total latency}, given by
\vs
\begin{equation}
{\cal L}(f) =  \sum_{e\in E}  f_e \cdot \ell_e(f_e)=  \sum\limits_{p\in \pset}  f_p \cdot \ell_p(f), \label{eq:totlatfpath}
\vs
\end{equation}
where $\ell_p(f) = \sum_{e\in p}\ell_e(f_e)$ denotes the latency on path $p$.
We denote the flow that minimizes the total latency by
\vs
\begin{equation} 
f^* \in \underset{f {\rm \ is \ feasible}}{\rm arg min} \ {\cal L}(f). 
\vs
\end{equation}

A \emph{routing problem} is given by $G=\left(V,E,\cee,\left\{\ell_e\right\}\right)$.
Classes of routing problems are denoted by $\gee$.
For $d\geq1$, we write $\gee^d$ to denote the class of all routing problems with latency functions of the form $\ell_e(f_e) = a_e(f_e)^d + b_e$, where $a_e,b_e\geq0$ are edge-specific constants.

To study the effect of slowdown information on self-interested behavior, this paper models the above routing problem as a heterogeneous non-atomic congestion game.
The slowdown-sensitivities of the users in commodity $c$ are modeled by a monotone, nondecreasing function $\beta^c:[0,r^c]\rightarrow [0,1]$, where each user $x \in [0,r^c]$ has a slowdown sensitivity $\beta^{c,x} \in [0,1]$.
Here, if user $x$ has $\beta^{c,x}=0$, this indicates that the user is purely delay-averse and is unresponsive to slowdowns.
On the other hand, if user $x$ has $\beta^{c,x}=1$, this indicates that the user cares nothing for \emph{actual} delay, and selects routes solely on the basis of how overcongested the route is compared to nominal.
The analysis in this paper assumes that each sensitivity distribution function $\beta^c$ is unknown to the system operator, and we write $\bee$ to denote the set of all feasible sensitivity distribution functions

The system operator provides all users with information regarding the slowdown experienced on the various paths; this is modeled by a number $\sig\in[0,1]$.
When $\sig=1$, users are provided complete and true information regarding each road's slowdown; when $\sig=0$, users are provided no information regarding slowdowns.
Altogether, given a flow $f$, the subjective cost that user $x\in[0,r^c]$ experiences for using path ${p} \in \pset^c$ is of the form%
%
%
\begin{align}
J^{c,x}(p;f) 	&= \sum\limits_{e\in {p}}\left[\left(1-\sig\beta^{c,x}\right)\ell_{e}(f_{e}) + \sig\beta^{c,x}\left(\ell_e(f_e)-\ell_e(0)\right)\right] \nonumber \\
					&= \sum\limits_{e\in {p}}\left[\ell_{e}(f_{e}) - \sig\beta^{c,x}\ell_e(0)\right]. \label{eq:costs}
\end{align}
Thus, each user $x\in[0,r^c]$ can be viewed as interpreting the cost of a road as the difference between its latency and a moderately-scaled version of its free-flow delay.
We assume that each user selects the lowest-cost path from the available source-destination paths.
We call a flow $f$ a \emph{Nash flow} if all users are individually using minimum-cost paths given the choices of other users.
That is, for all commodities $c\in\cee$, every user $x \in [0,r^c]$ using path $p$ in $f$ experiences a cost satisfying
\begin{equation}
J^{c,x}(p;f) = \min_{\tilde{p} \in \pset^c} J^{c,x} \left(\tilde{p};f\right).
\end{equation}
%
The above game is an \emph{exact potential game} with a convex potential function (see Lemma~\ref{lem:potential}), which implies that its set of Nash flows is nonempty and convex.

.

\subsection{Avoiding perverse signaling}

As an initial step towards characterizing a system operator\rq{}s optimal signaling policy, this paper compares the equilibrium total latency resulting from slowdown signaling with the un-influenced equilibrium total latency.
One simple measure which captures the possible \emph{harm} of slowdown signaling is the \emph{perversity index} as introduced in~\cite{Brown2020b}:
\begin{equation} \label{eq:pervdef}
\perv\left(\gee,\sig\right) := \sup_{G\in\gee}\sup_{\beta\in\mathcal{B}} \frac{\Lnf(G,\beta,\sig)}{\Lnf(G,\beta,0)}.
\end{equation}

Here, if a class of routing problems $\gee$ and signal parameter $\gamma$ have a large perversity index, this indicates that there exist routing problems in $\gee$ for which it would be better for the system planner to avoid signaling altogether.
If $\perv(\gee,\sig)>1$ for some $\gamma$, we say that $\gamma$ is a \emph{perverse} signaling policy.
If $\perv(\gee,\sig)=1$, we say that $\gamma$ is \emph{non-perverse.}

\section{Related Work} \label{sec:background}

Since the seminal work on price of anarchy for nonatomic routing games~\cite{Roughgarden2005}, these games have provided fertile ground to investigate a wide range of questions.
A common theme in these involves augmenting users' costs in some way to investigate the effect of some external influence or internal bias.
For example, the altruism models of~\cite{Chen2014,Brown2020a} assume that each member $x\in[0,1]$ of the population has an ``altruism" parameter $\alpha^x\in[0,1]$ which enters their edge cost function in the following way:
\begin{equation}
J_e^x(f_e) = \ell_e(f_e) + \alpha^xf_e\ell'_e(f_e).
\end{equation}
Here, $\alpha^x=1$ represents a fully altruistic user whose singular goal is to reduce aggregate congestion, and $\alpha^x=0$ represents a fully selfish user whose singular goal is to minimize personal travel time.
This model of altruism has been studied extensively, and much is known about the effects of this type of altruistic bias --- notably, in many situations, increased levels of altruism lead to reductions in equilibrium delays.
Many other similar cost function biases have been studied which take a similar form of $\ell_e(f_e)+b_e(f_e)$ for some specified $b_e(f_e)$, including pessimism~\cite{Meir2015b}, risk-aversion~\cite{Lianeas2016}, and various generalizations~\cite{Kleer2017,Meir2018}.

The above work is concerned with \emph{characterizing} the effect of user preferences on equilibrium behavior.
In parallel, many researchers have investigated methods of \emph{influencing} user preferences to improve equilibrium behavior.
These works on influence often involve similar modifications of user costs.
For example, pricing~\cite{Ferguson2021} is a common means of influencing user choices, and a popular pricing model allows the system planner to assign pricing functions to each edge of $\tau_e(f_e)$, and assumes (similar to the altruism model above) that users have heterogeneous price-sensitivities $s^x$:
\begin{equation}
J_e^x(f_e) = \ell_e(f_e) + s^x\tau_e(f_e).
\end{equation}
This is clearly tightly connected with the altruism model, particularly if the $\{\tau_e\}$ are chosen to be marginal-cost prices of the form $\tau^{\rm mc}_e(f_e) = f_e\ell'_e(f_e)$.
This has led to various synergies between the study of altruism and marginal-cost pricing~\cite{Brown2020a}, as results in one area neatly imply conclusions in the other.

\section{Our Contributions on Slowdown Signaling} \label{sec:ourContributions}

\subsection{Providing slowdown information can harm congestion on some networks}

Our first result shows that no universal signaling policy exists: networks exist for which all nonzero slowdown signals can increase aggregate congestion costs.
However, some classes of networks  are immune to these pathologies; in particular, slowdown signaling has the potential to improve congestion on all parallel networks.
In the following, for simplicity of exposition, each edge\rq{}s latency function is assumed to be a polynomial of the form $\ell_e(f_e) = a_e(f_e)^d + b_e$, for edge-specific nonnegative coefficients $a_e$ and $b_e$.

\begin{theorem} \label{thm:one}
Let $\gee$ be the class of all routing problems.
For all $\sig\in(0,1]$, routing problems exist for which $\sig$ is a perverse signal:
\begin{equation}
\perv\left(\gee,\gamma\right) >1. \label{eq:thm perv}
\end{equation}
However, let $\gee_p^d$ be the class of all parallel routing problems with latency functions of the form $\ell_e(f_e) = a_e(f_e)^d + b_e$.
Then non-perverse signaling is possible.
In particular,
\begin{equation}
\sig\in[0,d/(d+1)] \quad\quad\mbox{if and only if}\quad\quad \perv\left(\gee_p^d,\gamma\right) = 1. \label{eq:thm nonperv}
\end{equation}
\end{theorem}
\vspace{2mm}

Note that since $\gee_p^d\subset \gee$, the results in~\eqref{eq:thm perv} and~\eqref{eq:thm nonperv} indicate that there exist networks for which all signaling policies are perverse, but that these pathological networks are \emph{never} parallel networks.
Thus, a system planner who knows they are working with parallel networks can safely employ any signal satisfying $\sig\in[0,d/(d+1)]$ without fear of causing harm relative to the uninfluenced equilibria.

The proof of this theorem relies on Lemma~\ref{lem:tollequiv}, which explicitly relates the equilibrium flows under slowdown signaling to a similar formulation which is reminiscent of user costs experienced under marginal-cost tolls.
This allows us to leverage existing results on marginal-cost tolls to obtain the proof of the theorem.

\begin{lemma} \label{lem:tollequiv}
Let $\beta\in\bee$, $G\in\gee^d$, $\gamma\in[0,1)$. 
For each $x\in[0,1]$, let 
\begin{equation}\label{eq:alpha}
\alpha^x:= \frac{\sig\beta^x}{d(1-\sig\beta^x)}.
\end{equation}
Then the set of Nash flows associated with the game $(G,\beta,\gamma)$ is equal to the set of Nash flows for a game with the same network and a user-specific edge cost function of
\begin{equation}\label{eq:emulated}\vs
J^x_e(f_e) = (1+d\alpha^x)a_e(f_e)^d + b_e.
\end{equation}\vspace{.5mm}
\end{lemma}

The proof of Lemma~\ref{lem:tollequiv} is provided in the Appendix.
The cost functions~\eqref{eq:emulated} in Lemma~\ref{lem:tollequiv} are well-studied in the literature on heterogeneous nonatomic routing games.
These cost functions are identical to those induced by marginal-cost tolls when users are heterogeneous in price sensitivity; they also appear in the $\alpha$-altruism model of~\cite{Chen2014,Brown2020a}.
Thus, results from those two streams of work may be leveraged to draw conclusions about the effects of slowdown signaling.

\subsubsection*{Proof of Theorem~\ref{thm:one}}
Due to Lemma~\ref{lem:tollequiv}, the proof of Theorem~\ref{thm:one} follows in a straightforward manner from Theorem~(2) of~\cite{Brown2020b} and Theorem~7.1 of~\cite{Chen2014}.
\hfill\qed

\subsection{Optimal signaling need not be truthful}

Our next result considers the goal of computing the optimal signaling policy which minimizes worst-case equilibrium traffic congestion.
%
%
Here, we consider a situation in which the planner possesses additional information and in which case aggressive signaling mechanisms may be warranted.
One example of this is reminiscent of the setting studied in~\cite{Brown2017c}: the planner knows \emph{a priori} that the traffic rate on the network is high enough that all edges would be used in an un-influenced Nash flow.
Here, suppose that the planner knows that for every user $x\in[0,1]$, the slowdown-sensitivity satisfies $\beta^x\in[\bl,\bu]$, where $\bl>0$ and $\bu<1$.
In this setting, for linear-affine-cost parallel networks, the planner's optimal signal may actually \emph{over-state} the network's true slowdowns:
\begin{theorem}\label{thm:aggressive}
Let $\bar{\gee}_p^1$ denote the class of linear-latency parallel networks in which every edge has positive traffic in an un-influenced Nash flow.
For every $G\in\bar{\gee}_p^1$, it holds that
\begin{equation}\label{eq:fugam}
\sig^*:=\frac{1}{\bl+\bu} = \argmin_{\sig\geq0} \max_{\beta} \Lnash\left(G,\beta,\sig\right).
\end{equation}
Under the influence of this signal, the worst-case equilibrium total latency satisfies
\begin{equation}\label{eq:poafu}
\max_{G\in\bar{\gee}_p^1}\max_{\beta} \frac{\Lnash\left(G,\beta,\sig^*\right)}{\Lopt(G)} = \frac{4}{3}\left(1-\frac{\bl/\bu}{\left(1+\bl/\bu\right)^2}\right).
\end{equation}
\end{theorem}

\noindent The proof of Theorem~\ref{thm:aggressive} is provided in the Appendix.

\subsection{Broader Implications for Heterogeneous Congestion Games}

An interesting byproduct of the analysis required for this work is that we have discovered a weighted potential game formulation which applies to a large class of \emph{heterogeneous} nonatomic congestion games; to the best of our knowledge, this formulation is novel.
This formulation opens the door to computing Nash flows efficiently for a large class of games by performing gradient descent on the potential function of an associated exact potential game.
To understand the new formulation, first consider the following lemma which demonstrates that the slowdown-sensitivity games in this paper are exact potential games:
\vspace{2mm}
\begin{lemma} \label{lem:potential}
For any $\beta\in\bee$, $G\in\gee$, and $\gamma\geq0$, the game $(G,\beta,\gamma)$ specified by cost functions~\eqref{eq:costs} is an exact potential game with a convex potential function.
Thus:
\begin{enumerate}
\item $(G,\beta,\gamma)$ has at least one Nash flow, 
\item the set of Nash flows is convex, and
\item for each user $x\in[0,1]$, given two Nash flows $f$ and $f\rq{}$ (in which user $x$ chooses paths $p$ and $p\rq{}$, respectively), it holds that $J^x(p;f)=J^x(p\rq{};f\rq{}).$
\end{enumerate}
\end{lemma}
\noindent The proof of Lemma~\ref{lem:potential} is provided in the Appendix.
We can now state the main result, that many heterogeneous nonatomic congestion games with a particular form of polynomial cost function are weighted potential games and that their Nash flows exactly coincide with those of our slowdown-sensitivity games.
\begin{theorem}\label{thm:wpg}
In some routing problem $G\in\gee^d$, let every user $x\in[0,1]$ have altruism parameter $\alpha^x\in[0,1]$ and experience a cost on edge $e$ of $J^x_e(f) = (1+d\alpha^x)a_e(f_e)^d + b_e.$
Then routing problem $G$ coupled with the user population described by $\alpha^x$ is a weighted potential game whose set of Nash flows is equal to that of slowdown-sensitivity game $(G,\beta,1)$, where for each $x\in[0,1]$ it holds that
\begin{equation}\label{eq:beta}
\beta^x = \frac{\alpha^xd}{\alpha^xd +1}.
\end{equation}
\end{theorem}

\noindent The proof of Theorem~\ref{thm:wpg} is provided in the Appendix.

\section{Conclusion}
This paper explores the connections between a new class of slowdown-sensitive congestion games and established results for heterogeneous nonatomic routing games under the influence of altruism and marginal-cost pricing.
In particular, we demonstrate a tight connection between the two, and illustrate how to connect results from one to the other.
Finally, we exploit this connection to demonstrate a novel result for certain heterogeneous nonatomic congestion games, namely that they are weighted potential games and thus their equilibria may be computed efficiently.


\bibliographystyle{splncs04}
\bibliography{../../library/library}

\section*{Appendix: Proofs}

\subsubsection*{Proof of Lemma~\ref{lem:tollequiv}.}
By~\eqref{eq:alpha}, emulated cost functions~\eqref{eq:emulated} can be written
\begin{equation} \label{eq:emulated1}
J^x_e(f_e) = \frac{a_e(f_e)^d}{1-\sig\beta^x} + b_e.
\end{equation}
Since users\rq{} ordinal preferences (and thus Nash flows) are invariant to multiplication by user-specific constants, the cost functions in~\eqref{eq:emulated1} are equivalent to ones given by
\begin{equation} \label{eq:emulated2}
\tilde{J}^x_e(f_e) = a_e(f_e)^d + (1-\sig\beta^x)b_e.
\end{equation}
Evidently, the cost functions~\eqref{eq:emulated2} are equal to those of the nominal game $(G,\beta,\sig)$ (see~\eqref{eq:costs}).
Since, for each user $x$, these cost functions encode the same ordinal preferences as those given by~\eqref{eq:emulated}, the two sets of cost functions induce identical sets of Nash flows.\hfill\qed

\subsubsection*{Proof of Theorem~\ref{thm:aggressive}.}
The optimal signal factor~\eqref{eq:fugam} can be deduced from the results in~\cite[Theorem 1]{Brown2017c} regarding the optimal marginal-cost-toll scale factor in an identical setting.
Our Lemma~\ref{lem:tollequiv} provides that once a value of $\gamma$ is fixed, each slowdown-sensitivity $\beta^x$ distribution can be used to compute a taxation-sensitivity $\alpha^x$ distribution which induces an identical set of Nash flows.
In~\cite[Theorem 1]{Brown2017c}, it is shown that worst-case congestion is minimized for scaled marginal-cost taxes when the taxation-sensitivity distribution satisfies
\begin{equation} \label{eq:fuequation}
\max_x\alpha^x = \frac{1}{\min_x\alpha^x}.
\end{equation}
Thus, due to the equivalence between slowdown-sensitivity and taxation sensitivity and applying the transformation~\eqref{eq:alpha} from Lemma~\ref{lem:tollequiv}, the optimal signal factor $\sig^*$ satisfies
\begin{equation}
\frac{\sig^*\bl}{1-\sig^*\bl} = \frac{\sig^*\bu}{1-\sig^*\bu},
\end{equation}
which implies~\eqref{eq:fugam}.
The upper bound~\eqref{eq:poafu} follows immediately from~\cite[Lemma 3.1]{Brown2016a}.
\hfill\qed

\subsection*{Proof of Lemma~\ref{lem:potential}.}
Note that the individual slowdown sensitive cost function~\eqref{eq:costs} is a sum of two terms: $\ell_e(f_e)$ and $\sig\beta^x \ell_e(0)$.
A game with costs defined by only the first term is simply a homogeneous nonatomic routing game, which is well-known to be an exact potential game with a convex potential function~\cite{Roughgarden2005,Sandholm2009}.
A game with costs defined by only the second term is a trivial game in which each user\rq{}s cost function depends only on her own action; any such game is also known to be an exact potential game with a linear potential function~\cite{Sandholm2009}.
In~\cite{Sandholm2009} it is shown that a game whose cost functions are the sum of exact potential games\rq{} cost functions is itself an exact potential game whose potential function is the sum of the potential functions of its component games.
Therefore, $(G,\beta,\gamma)$ is an exact potential game with a convex potential function --- and this convexity implies points (1), (2), and (3).\hfill\qed

\subsection*{Proof of Theorem~\ref{thm:wpg}.}
The proof relies on a serial application of Lemmas~\ref{lem:tollequiv} and~\ref{lem:potential}.
In particular, Lemma~\ref{lem:tollequiv} provides a bijection between the heterogeneous altruistic congestion games assumed by Theorem~\ref{thm:wpg} and the slowdown-sensitivity games considered in this paper.
Equation~\eqref{eq:beta} is simply the inverse mapping of~\eqref{eq:alpha}; thus, the cost functions of the two games express identical ordinal preferences and are related by a set of user-specific constant multipliers.
Thus, the games are (for the purposes of equilibrium computation) equivalent: equilibria computed for one are automatically equilibria for the other.
Accordingly, in light of Lemma~\ref{lem:potential}, one may use potential game equilibrium-finding techniques (e.g., gradient descent) to compute the equilibria of the slowdown-sensitivity game and thereby find the equilibria of the altruistic congestion game.
\hfill\qed


\end{document}